\documentclass[12pt]{article}%
\usepackage{graphicx}

\begin{document}
 
\begin{center}{\LARGE\bf Main problems in constructing quantum theory based on finite mathematics} \end{center}

\vskip 1em \begin{center} {\large Felix M. Lev} \end{center}

\begin{center} Email:  felixlev314@gmail.com \end{center}
\begin{abstract}
As shown in our publications, quantum theory based on a finite ring of characteristic $p$ (FQT) is more general than standard quantum theory (SQT)  because the latter is a degenerate case of the former in the formal limit $p\to\infty$. One of the main differences between SQT and FQT is the following. In SQT, elementary objects are described by irreducible representations (IRs) of a symmetry algebra in which energies are either only positive or only negative and there are no IRs where there are states with different signs of energy. In the first case, objects are called particles, and in the second - antiparticles. As a consequence, in SQT it is possible to introduce conserved quantum numbers (electric charge, baryon number, etc.) so that particles and antiparticles differ in the signs of these numbers. However, in FQT, all IRs necessarily contain states with both signs of energy. The symmetry in FQT is higher than the symmetry in SQT because one IR in FQT splits into two IRs in SQT with positive and negative energies at $p\to\infty$. Consequently, most fundamental quantum theory will not contain the concepts of particle-antiparticle and additive quantum numbers. These concepts are only good approximations at present since at this stage of the universe the value $p$ is very large but it was not so large at earlier stages. The above properties of IRs in SQT and FQT have been discussed in our publications with detailed technical proofs. The purpose of this paper is to consider models where these properties
can be derived in a much simpler way.
\end{abstract}

\begin{flushleft} Keywords: finite mathematics; standard mathematics; finite quantum theory; standard quantum theory\end{flushleft}
\begin{flushleft} MSC 2020: 11Axx, 11Txx, 13Mxx, 16Gxx, 81R05\end{flushleft}

\begin{center} {\bf  List of Abbreviations} \end{center}
\begin{flushleft} FM: finite mathematics\end{flushleft}
\begin{flushleft} SM: standard mathematics\end{flushleft}
\begin{flushleft} SR: special relativity \end{flushleft}
\begin{flushleft} NM: nonrelativistic mechanics\end{flushleft}
\begin{flushleft} QT: quantum theory\end{flushleft}
\begin{flushleft} CT: classical theory\end{flushleft}
\begin{flushleft}  FQT: Quantum theory based on finite mathematics \end{flushleft}
\begin{flushleft}  SQT: Standard quantum theory \end{flushleft}
\begin{flushleft} IR: irreducible representation\end{flushleft}
\begin{flushleft}  QFT: Quantum Field Theory\end{flushleft}
\begin{flushleft}  NQT: Nonrelativistic Quantum Theory\end{flushleft}
\begin{flushleft}  RQT: Relativistic Quantum Theory\end{flushleft}
\begin{flushleft}  dS: de Sitter \end{flushleft}
\begin{flushleft}  AdS: Anti de Sitter \end{flushleft}
\begin{flushleft}  dSQT: de Sitter Quantum Theory \end{flushleft}
\begin{flushleft}  AdSQT: Anti de Sitter Quantum Theory \end{flushleft}

\section{The main goal of this paper}
\label{goal}

One of the key problems of
QFT is the problem of divergences: the theory gives divergent expressions
for the S-matrix. While in renormalized theories, the divergences can be eliminated by
renormalization, in non-renormalized QFTs, they cannot be eliminated and this is a great obstacle for constructing quantum gravity based on QFT.

The problem of divergences has been considered by many physicists, and
there has long been an idea in the air that this problem can only be solved within the framework of a discrete and finite quantum theory. It would seem natural to think that such a theory should proceed from discrete and finite mathematics. 
However, most mathematicians and physicists believe that SM (with infinities and continuities) is fundamental while discrete and finite mathematics is a science of a lower rank which is only needed for applications in some models. 
This point of view has developed for historical reasons (because more than 300 years ago Newton and Leibniz proposed the calculus of infinitesimals) and due to the fact that SM has achieved many impressive successes in describing experimental data. 

The calculus of infinitesimals seemed natural when people did not know about elementary particles and thought that any substance could be divided into any arbitrarily large number of arbitrarily small objects.
But now we know that at the level of elementary particles  there are no arbitrarily small parts and no continuity.

Also, history tells us that if a theory successfully describes many experimental data, this is not yet a guarantee that this theory is the most fundamental. For example, NM successfully describes a lot of experimental data and before the creation of SR it was believed that NM was a fundamental theory. SR did not refute NM, but showed that the latter is a
degenerate case of the former in formal limit $c\to\infty$ where $c$ is usually treated as
the speed of light. As shown in our works \cite{book,arxiv,lev3,FM}, FM is more general (fundamental) than SM: the latter is a degenerate case of the former in formal limit $p\to\infty$ where $p$ is a characteristic of a ring in FM.

Several famous physicists (e.g., Gross, Nambu, Schwinger and Weyl) discussed approaches when QT involves FM (see e.g., \cite{Nambu}). They are called hybrid quantum systems and described in \cite{Vourdas}. The reason is that here physical quantities
belong to a finite ring but quantum states are elements of standard Hilbert spaces.
On the other hand, in \cite{book,arxiv,lev3}, we have proposed an approach called finite quantum theory (FQT) where not only physical quantities but also quantum states are described by finite rings. We have shown that FQT is more general (fundamental) than SQT: SQT is a degenerate case of FQT in formal limit $p\to\infty$ where $p$ is the characteristic of the ring in FQT.

In SQT, elementary objects are described by IRs of symmetry algebras in which energies can be either $\geq 0$  or $\leq 0$ and there are no IRs with both positive and negative energies. In the first case, objects are called particles and in the second - antiparticles, and after secondary quantization, the energies of antiparticles also become positive. In SQT, particles and antiparticles are characterized by additive quantum number, e.g., the electric charge, the baryon number and others. If particle A is characterized by some additive quantum numbers and antiparticle B has the same mass and spin as A, but additive quantum numbers of B 
are equal to the corresponding additive quantum numbers of A with the opposite sign, then B is called the antiparticle for A. In SQT there are superselection rules that prohibit the superposition of a particle and its antiparticle. For example, electron-positron or proton-antiproton superpositions are prohibited,
and this is interpreted as a consequence of the conservation of electric charge and baryon quantum number.

However, in FQT, one IR necessarily contains states with both positive and negative energies. Since such states belong to the same IR, their superpositions are allowed and there are no superselection rules. One can {\it formally} call states with positive energies particles and assign some additive quantum numbers to them, and call states with negative energies antiparticles and assign opposite quantum numbers to them. Then it turns out that there are no conservation laws for such quantum numbers, and, for example, electron-positron or proton-antiproton superpositions are allowed. It is clear that this completely contradicts the basic concepts of SQT. In other words, in FQT, the standard concepts of electric charge, baryon quantum number and other additive quantum numbers do not work.

This situation may prompt physicists to declare that FQT is not a physical theory and should be rejected. However, symmetry in FQT is higher than symmetry in SQT since one IR in FQT splits into two IRs in SQT with positive and negative energies in the formal limit $p\to\infty$ when FM goes to SM. At the popular level, this situation can be described as follows.

Suppose there are two theories, Theory 1 and Theory 2. In Theory 1, energies in IRs are represented by points on a circle so that the energies on the right semicircle are called positive, and on the left semicircle negative. Since states with positive and negative energies belong to the same IR, their superpositions are allowed.
Now let's suppose that in Theory 2 there are two types of IRs: in IRs of the first type, energies can only be positive, and in IRs of the second type - only negative. Then superpositions of states with positive and negative energies are prohibited since such states belong to different IRs. Then Theory 1 in which there is one IR describing the entire circle has higher symmetry than Theory 2 in which there are two IRs describing the right and left semicircles independently.

In such a scenario, the fact that at the present stage of the evolution of the universe, SQT describes experiments with very high accuracy follows the fact that at this
stage, the quantity $p$ is very large. As shown in \cite{book,arxiv}, within the framework of semiclassical approximation to FQT, it is possible to derive the law of universal
gravitation where the gravitational constant $G$ is proportional
to $1/ln(p)$. By comparing this result with the experimental
value of $G$, one gets that $ln(p)$ is of the order of $10^{80}$ or more, and therefore $p$ is
a huge number of the order of $exp(10^{80})$ or more. However, $p$ cannot be treated as an infinite number because, since $ln(p)$ is "only" of
the order of $10^{80}$, gravity is observable. At the same time, in \cite{book,arxiv} we have
made arguments that in early stages of the universe, the value of $p$ was much smaller than
now and so in these stages only FQT may be reliable for describing different experimental
data.  
In \cite{book,arxiv} we considered several other phenomena where it is important that
$p$ is finite and not infinitely large. 

There is an analogy here with the fact that when speeds are much less than $c$
one can consider $c$ infinitely large and then NM describes these phenomena with great
accuracy. However, when speeds are comparable to $c$, $c$ cannot be considered an
infinitely large value and then only SR can be reliable.

These remarks indicate that the construction of a fundamental quantum theory based on a finite
$p$ will be a problem based on fundamentally new concepts: since the concepts of
 particle-antiparticle, electric charge and baryon quantum number have a physical meaning only
for very large values of $p$, then in such a theory, in the most general case, there should 
be no such concepts. However, from the point of view of the development of science, the
fundamental quantum theory at finite $p$ must be constructed. 

When we compare two theories A and B, a question arises what criteria should be used to prove that, for example, theory A is more general than theory B and B is a degenerate case of A.
In \cite{book,arxiv,lev3} we have proposed the following criteria

{\bf Definition:} {\it Let theory A contain a finite nonzero parameter and theory B be obtained from theory A in the formal limit when the parameter goes to zero or infinity. Suppose that, with any desired accuracy, A can reproduce any result of B by choosing a value of the parameter. On the contrary, when, the limit is already taken, one cannot return to A and reproduce all results of A. Then A is more general than B and B is a degenerate case of A}. 

The proofs in \cite{book,arxiv} of fundamental facts that FM is more general (fundamental) than SM and FQT is more general (fundamental) than SQT contain a lot of technical details and, as a result, these works are quite long (291 and 293 pages, respectively). In this paper, we present two simple examples that require a minimum of prior knowledge and which we hope will stimulate readers to explore these fundamental facts in more depth.

The paper is organized as follows. In Secs. \ref{finmath} and \ref{FQTvsSQT} we describe  basic facts about finite rings and quantum theory based on finite mathematics. In Secs. 
\ref{supersymmetry} and \ref{model} we describe supersymmetry in AdS theory and
consider a simple model example demonstrating the difference between SQT and FQT.
In Secs. \ref{Dsingl} and \ref{DsinglIRs} this difference is demonstrated on the
example of Dirac supersingletons.

\section{Basic facts about finite rings}
\label{finmath}

In contrast to SM which starts from the infinite ring  $Z=(-\infty,...-2,-1,0,1,2,...\infty)$, FM starts from the finite ring $R_p=(0, 1, 2, ... p-1)$ where addition, subtraction and multiplication are defined
as usual but modulo $p$. We believe that the notation $Z/p$ for $R_p$ is not adequate because it may
give a wrong impression that FM starts from the infinite set $Z$
and that $Z$ is more general than $R_p$. However, although $Z$ has more elements than $R_p$, $Z$ cannot 
be more general than $R_p$ because $Z$ does not contain operations modulo a number. If $p$ is prime then $R_p$ becomes
the Galois field $F_p$ but in this paper we  
consider only finite rings. The theory of such rings is described in textbooks (see e.g., \cite{VdW,Ireland,Davenport}). The number $p$ is called the characteristic
of the ring $R_p$. For example, if $p=5$ then 3+1=4 as usual but 3$\cdot$2=1, 4$\cdot$3=2, 4$\cdot$4=1 and 3+2=0. Therefore -2=3, -4=1 etc. 

One might say that the above examples have nothing to do with reality since 3+2 always equals 5 and not zero.
However, since operations in $R_p$ are modulo $p$, one can represent 
$R_p$ as a set $\{0,\pm 1,\pm 2,...,\pm(p-1)/2)\}$ if $p$ is odd or as a set
$\{0,\pm 1,\pm 2,...,\pm (p/2-1),p/2\}$ if $p$ is even. Let $f$ be a function from $R_p$ to $Z$ such that
$f(a)$ has the same notation in $Z$ as $a$ in $R_p$. 

\begin{figure}[!ht]
	\centerline{\scalebox{0.3}{\includegraphics{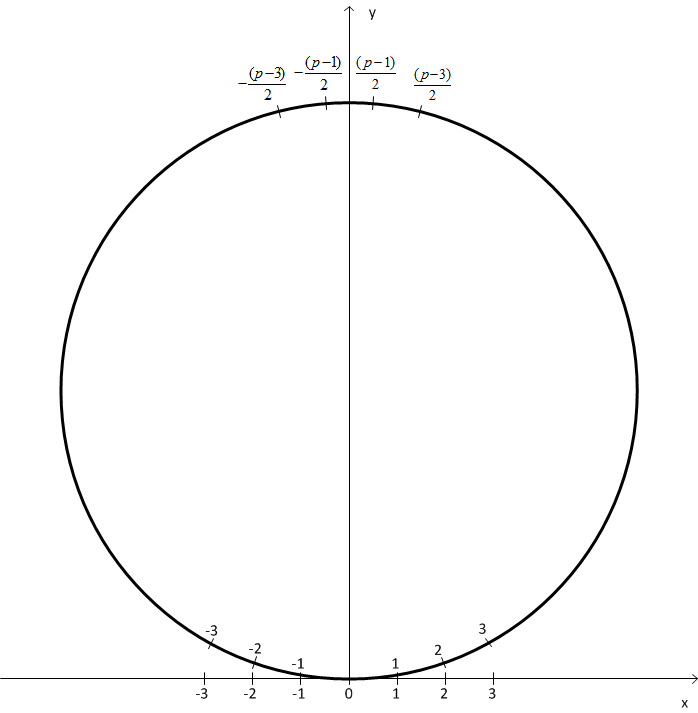}}}
	\caption{
		Relation between $R_p$ and $Z$
	}
	\label{Fig.2}
\end{figure}

If elements of $Z$ are depicted as integer points on the $x$ axis of the $xy$ plane then, if $p$ is odd, the elements of $R_p$
can be depicted  as points of the circumference in Figure \ref{Fig.2}
and analogously if $p$ is even.

Formally, we can call the element $a\in R_p$ positive if $f(a)>0$ and negative if $f(a)<0$. In other words, the element $a\in R_p$ is positive if it is in the right half-plane of  Figure 1 and negative if in the left half-plane. While in SM, a sum of two positive numbers is always positive and greater than both original numbers, in FM 
(where calculations are carried out modulo $p$), it is even possible that a sum of two positive numbers is negative. For example,
$(p-1)/2+1=(p+1)/2=-(p-1)/2$. However, for numbers $a$ such that $|f(a)|$ is much less than $p$, the results of all the operations are the same as in $Z$, i.e., for such numbers we do not notice the existence of $p$. 

When $p\to\infty$, a vicinity of zero in $R_p$ becomes the infinite set $Z$. 
Therefore {\it even from pure mathematical point of view}, the concept of infinity cannot be fundamental because, as soon as we replace $R_p$ by $Z$, we automatically obtain a degenerate theory because in $Z$ there are no operations modulo a number.

In FQT, states are elements of linear spaces over $R_p$. One might think that SQT is more general than FQT because in SQT one can work not only with integers but also with rational and real numbers. However, as noted 
in \cite{book,arxiv,lev3,FM} and Sec. \ref{FQTvsSQT}, since in SQT the states are projective, for describing wave functions with any desired accuracy it suffices to use only integers.

\section{Do we need spacetime background in quantum theory?}
\label{background}

Historically, the concepts of background space and fields in this space arose from classical electrodynamics and then they were further developed in General Relativity which is also a classical (i.e., non-quantum) theory. For example, now we know that the electromagnetic field consists of photons but, at the classical
level, the theory does not describe the state of each photon. The classical electromagnetic
fields ${\bf E}({\bf r}, t)$ and ${\bf B}({\bf r}, t)$ describe the {\it effective} contribution of all photons at
the point $x = ({\bf r}, t)$ of Minkowski space, and in classical (non-quantum) theory it is assumed that the parameters $x = ({\bf r}, t)$ can be measured with any desired accuracy. 
This is similar to the situation in statistical physics, where systems of many particles are considered, but the theory does not describe each particle individually, but introduces concepts that make sense only for ensembles of a large number of particles 
(for example, temperature and pressure).

In quantum theory, every physical quantity must be described by an operator. For example, one can talk about coordinates of a certain particle only if the position operator for this particle is defined. In QFT, particles are described by field operators
$\Psi(x)$ where $x$ is a point in the background space (e.g., in Minkowski space). 
When there are many particles, it may seem that they are in some space. However, such space is only a mathematical and not a physical object.
It is not related to
certain particles  and there are no operators for the coordinates in this space.
The goal of QFT is to construct the S-matrix and when the theory is already constructed one
can forget about Minkowski space because $x$ is only an integration parameter in the expression for the S-matrix and no physical quantity depends on $x$. This
is in the spirit of the Heisenberg S-matrix program according to which in RQT it is impossible to describe the state of the system at each moment in time and it is possible to describe only transitions of states from the infinite
past when $t\to -\infty$ to the distant future when $t\to +\infty$.

Note that the fact that the S-matrix is the operator in momentum space
does not exclude a possibility that in some situations it is possible to have a spacetime
description with some accuracy but not with absolute accuracy. First of all,
as noted by Pauli \cite{Pauli}, the problem of time is one of the most important unsolved problems of quantum theory because there is no time operator. Also, the position operator
in momentum representation usually exists not only in nonrelativistic theory but in
relativistic theory as well. In this case it is known as the Newton-
Wigner position operator \cite{NW} or its modification. However, the coordinate description of elementary
particles can be only approximate. For example, coordinates of a particle with the mass $m$ cannot be measured
with the accuracy better than the particle Compton wave length $\hbar/(mc)$ \cite{BLP}.

As noted in the extensive literature on QFT (see, e.g., \cite{Bog}), the use of field functions $\Psi(x)$ also leads to the following mathematical problem. 
Quantum interacting local fields can be treated only as operator distributions. A known fact from the theory of distributions is that their product at the same point is not a correct mathematical operation. Hence if $\Psi_1(x)$ and $\Psi_2(x)$ are two local operator fields then the product $\Psi_1(x)\Psi_2(x)$ is not well defined.  Physicists often ignore this problem: they think that such products are needed to preserve locality (although the operator of
the quantity $x$ does not exist). As a consequence, representation operators of
interacting systems constructed in QFT are not well defined and the theory contains
anomalies and infinities. A detailed discussion of other problems of QFT can be found, for example, in \cite{Concepts}.

Let us also note that so far the approaches to the spacetime background come from SM in which, as is known from Gödel's incompleteness theorems and other results, there are foundational problems. There is an extensive literature which conjectures that foundational
problem of quantum theory will be solved in the framework of approaches with fundamental length (see e.g., \cite{FLen} and references therein) because here it will
be possible to circumvent infinitesimals. However, since this literature is based on SM,
it involves infinitesimals implicitly.

Despite the noted mathematical problems, QFT is very popular among physicists due to the following. In renormalizable theories, divergences can be eliminated and, although from a mathematical point of view, renormalization (when operations with singularities yield non-singular expressions) is not a correct mathematical operation, 
in some cases it leads to very impressive agreements between theory and experiment.
However, in non-renormalizable theories, singularities cannot be eliminated, and this is the main obstacle to the construction of quantum gravity.

\section{Quantum theory based on finite mathematics}
\label{FQTvsSQT}

In this section, following \cite{book,arxiv,FM}, we briefly describe why SQT is a degenerate case of FQT in the limit $p\to\infty$.

In SQT, physical states are described by elements of complex Hilbert spaces, and operators of physical quantities are self-adjoint operators in such spaces. By analogy, in FQT, physical states are elements of linear spaces over the ring $R_{p^2}$ which is a quadratic generalization of $R_p$ and contains $p^2$ elements: any element of $R_{p^2}$ can be represented as $a+bi$ where $a,b\in R_p$ and $i$ is a formal element such that $i^2=-1$. Then the definition of addition, subtraction and multiplication in $R_{p^2}$ is obvious and $R_{p^2}$ is a ring regardless whether $p$ is prime or not. 

In both, SQT and FQT, dS symmetry is defined by the operators 
$M^{ab}$ ($a,b=0,1,2,3,4$, $M^{ab}=-M^{ba}$) satisfying the
commutation relations
\begin{equation}
[M^{ab},M^{cd}]=-i (\eta^{ac}M^{bd}+\eta^{bd}M^{ac}-\eta^{ad}M^{bc}-\eta^{bc}M^{ad})
\label{CR}
\end{equation}
where $\eta^{ab}$ is the diagonal tensor such that
$\eta^{00}=-\eta^{11}=-\eta^{22}=-\eta^{33}=-\eta^{44}=1$. As discussed in detail in \cite{book,arxiv,FM,PRD}, this definition does not 
involve the fact that the dS group is the group of motions of dS space.

The {\it definition} of AdS symmetry is given by the same expressions
but $\eta^{44}=1$. By analogy with the dS case, this definition does not 
involve the fact that the AdS group is the group of motions of AdS space.
At the same time, there is an extensive literature on AdS/CFT correspondence in which
AdS symmetry essentially involves the properties of classical AdS space 
(see e.g. \cite{AdSCFT1,AdSCFT2,AdSCFT3} and references therein). This literature yields interesting results but currently AdS/CFT correspondence is a conjectured relationship between two kinds of physical theories. As argued in \cite{book,arxiv,FM}, at the most fundamental level, quantum theory should not involve such classical concepts as AdS space and, in this paper, following \cite{book,arxiv,lev3} and other our publications, we argue  that quantum theory should be based on FM.

In SQT, operators
of physical quantities act in Hilbert spaces supplied by a scalar product (...,...), and these operators
are selfadjoint. In particular, the operators in Eqs. (\ref{CR}) are selfadjoint. 
However, in spaces over $R_{p^2}$ it is impossible to introduce a scalar product that satisfies the condition that $(x,x)>0$ for all non-zero elements $x$ in such spaces.
The matter is that, as explained in Sec. \ref{finmath}, in FM, the concepts $>$ and $<$ have their usual meaning only for those elements $a\in R_p$ for which $|f(a)|$ is much less than $p$. Therefore, in spaces over $R_{p^2}$ the concept of Hermitian conjugation has limited applicability: it makes sense only for the actions of operators on elements of spaces for which the expansion coefficients with respect to the basis elements are much less than $p$.

Let us define ${\tilde M}^{ab}=M^{ab}$ if $a,b\neq 4$ and ${\tilde M}^{ab}=iM^{ab}$ if
$a\neq 4,\,\,b=4$. Then a direct check shows that the set of operators ${\tilde M}^{ab}$ satisfies Eq. (\ref{CR})  if $\eta^{44}$ is replaced by $-\eta^{44}$. Therefore, if the set of operators $M^{ab}$ satisfies the conditions $(\ref{CR})$ for the dS algebra, then the set of operators ${\tilde M}^{ab}$ satisfies the conditions $(\ref{CR})$ for the AdS algebra and {\it vice versa}.

Therefore in FQT, the dS and AdS theories are equivalent. However, in SQT they are not
equivalent for the following reason. Here it is required that the operators $M^{ab}$ should not only satisfy Eq. (\ref{CR}) but additionally they should be selfadjoint (as explained above, in FQT such a requirement cannot be imposed). However, if the operators $M^{a4}$ are Hermitian then the operators ${\tilde M}^{a4}=iM^{ab}$ are anti-Hermitian.

When in SQT the operators in Eq. (\ref{CR}) are selfadjoint then, as described in a wide
literature, IRs of the dS and AdS algebras are infinite-dimensional.
Representations in spaces over a ring of nonzero characteristic are called
modular representations. According to the Zassenhaus theorem (see e.g., \cite{Zassenhaus,Jantzen}), all modular IRs are finite-dimensional. In \cite{lev1,lev2}
we constructed modular IRs of the algebras defined by Eq. (\ref{CR}).

In SQT, all Hilbert spaces are separable, i.e., they contain a countable dense subset. 
In such spaces it is always possible to choose a basis $(e_1,e_2,...e_n,...)$ such that the norm of each $e_j$ is an integer. The elements of such spaces can be denoted as 
$(c_1,c_2,...c_n,...)$ where all the expansion coefficients $c_j$ for such a basis  are complex numbers and can be represented as $c_j=a_j+ib_j$.
As explained in \cite{FM}, since spaces in quantum theory are projective, it follows from the results of the textbook \cite{Fomin} that: 

{\it In SQT, each element of a separable Hilbert space can be approximated with any desired accuracy by a finite linear combination 
\begin{equation}
	x=\sum_{j=1}^n c_je_j
	\label{lin}
\end{equation}	
where all the numbers $a_j$ and $b_j$ are integers, i.e., belong to $Z$.}

In FQT, quantum states also can be represented in the form (\ref{lin}) but here
the $c_j=a_j+ib_j$ are elements of $R_{p^2}$. As shown in \cite{book,arxiv,lev3,FM}, by using {\bf Definition} and the above results one can prove that FQT is more general (fundamental) than SQT and the latter is a special degenerate case of the former in the formal limit $p\to\infty$: when the numbers $(a_j,b_j)$ are such that $\forall j$, $|f(a_j)|$ and 
$|f(b_j)|$ are much less than $p$ then FQT reproduces all results of SQT but SQT cannot
reproduce all results of FQT if some of the numbers $(a_j,b_j)$ are comparable to $p$.

\section{Supersymmetry}
\label{supersymmetry}

In SQT, supersymmetry is valid in the AdS case but is not valid in the dS one. As shown in \cite{asymm}, in SQT, dS symmetry is more general than AdS one, and it may be a reason why supersymmetry has not been discovered yet. However, as shown in \cite{book,arxiv,lev3,FM}, SQT is a degenerate case of FQT in the formal limit $p\to\infty$, and, as shown in 
Sec. \ref{FQTvsSQT}, in FQT, dS and AdS symmetries are equivalent.

Representations of the osp(1,4) superalgebra
are described by 14 operators, as well as representations of the Poincare superalgebra.
However, effectively, representations of the osp(1,4) superalgebra can be described only by
four fermionic operators. The matter is that ten bosonic operators of the osp(1,4) 
superalgebra are the anticommutators of the four fermionic operators. This is
not the case for the Poincare superalgebra since the Poincare
algebra operators are obtained from the so(2,3) ones by
contraction. One can say that the representation of the
osp(1,4) superalgebra is the implementation of the idea that
supersymmetry is the extraction of the square root from the
usual symmetry (by analogy with the treatment of the
Dirac equation as a square root from the Klein-Gordon equation).

Let $(d_1',d_2',d_1'',d_2'')$ be the fermionic operators of the osp(1,4) superalgebra. 
They should satisfy the following
relations. If $(A,B,C)$ are any fermionic operators, [...,...]
is used to denote a commutator and $\{...,...\}$ to denote an
anticommutator then
\begin{equation}
	[A,\{ B,C\} ]=F(A,B)C + F(A,C)B
	\label{S30}
\end{equation}
where the form $F(A,B)$ is skew symmetric, $F(d_j',d_j")=1$
$(j=1,2)$ and the other independent values of $F(A,B)$ are
equal to zero. 

As shown by various authors (see e.g., \cite{book,arxiv,Heidenreich}), the operators $M^{ab}$ in 
Eqs. (\ref{CR}) can be expressed through bilinear combinations of the fermionic operators as follows:
\begin{eqnarray}
	&&h_1=\{d_1',d_1''\},\,\,h_2=\{d_2',d_2''\},\,\,M_{04}=h_1+h_2,\,\,
	M_{12}=L_z=h_1-h_2\nonumber\\
	&&L_+=\{d_2',d_1''\},\,\, L_-=\{d_1',d_2''\},\,\, M_{23}=L_x=L_++L_-\nonumber\\
	&&M_{31}=L_y=
	-i(L_+-L_-),\,\, M_{14}=(d_2'')^2+(d_2')^2-(d_1'')^2-(d_1')^2\nonumber\\
	&&M_{24}=i[(d_1'')^2+(d_2'')^2-(d_1')^2-(d_2')^2]\nonumber\\
	&& M_{34}=\{d_1',d_2'\}+\{d_1'',d_2''\},\,\,M_{30}=-i[\{d_1'',d_2''\}-\{d_1',d_2'\}]\nonumber\\
	&&M_{10}=i[(d_1'')^2-(d_1')^2-(d_2'')^2+(d_2')^2]\nonumber\\
	&&M_{20}=(d_1'')^2+(d_2'')^2+(d_1')^2+(d_2')^2
	\label{Mab}
\end{eqnarray}
where ${\bf L}=(L_x,L_y,L_z)$ is the standard operator of three-dimensional rotations.

We require
the existence of the generating vector $e_0$ satisfying the
conditions :
\begin{eqnarray}
	d_j'e_0=d_2'd_1''e_0=0, \quad d_j'd_j''e_0=q_je_0\quad (j=1,2)
	\label{S32A}
\end{eqnarray}
The full representation space can be obtained by successively
acting by the fermionic operators on $e_0$ and taking all
possible linear combinations of such vectors. The theory of self-adjoint IRs of the
osp(1,4) algebra has been developed by several authors (see e.g., \cite{Heidenreich}),
and in \cite{book,arxiv} this theory has been generalized to the case of FQT.

\section{Model example}
\label{model}

In this section we consider a simple model example when in Eq. (\ref{S32A}) there are only two
fermionic operators $(d',d")$ and one bosonic operator $h$ such that
\begin{equation}
	h=\{d',d''\},\quad [h,d']=-d',\quad [h,d'']=d''
	\label{osp12}
\end{equation} 
Here the first expression shows that the relations (\ref{osp12})
can be formulated only in terms of the fermionic operators. We will consider IRs of the superalgebra (\ref{osp12}) in SQT and FQT. 
\subsection{IRs of the superalgebra (\ref{osp12}) in SQT}
\label{SS1}
Consider an IR of the algebra (\ref{osp12}) generated by a vector $e_0$ such that
\begin{equation}
	d'e_0=0,\quad he_0=q_0e_0 , \quad (q_0>1/2)
	\label{e01}
\end{equation}
and define $e_n=(d'')^ne_0$ ($n=1,2,...$). Then $d'e_n=a(n)e_{n-1}$ where, as follows from Eq. (\ref{e01}), $a(0)=0,\,\,a(1)=q_0$
and 
\begin{equation}
	a(n)=q_0+n-1-a(n-1)
	\label{an1}
\end{equation} 
The solution of this equation is 
\begin{equation}
	a(n)=n/2+ (q_0-1/2)[1-(-1)^n]/2
	\label{anB}
\end{equation}
Therefore, $a(n) > 0\,\, \forall n$ and, as follows from Eq. (\ref{osp12}), 
$he_n=(n+q_0)e_n$. So, we have obtained an infinite-dimensional IR of the algebra (\ref{osp12}) with the basis $(e_0,e_1,e_2...)$ where all basis vectors are eigenvectors of the operator $h$ with positive eigenvalues. 

Consider now an IR of the algebra (\ref{osp12}) generated by a vector $f_0$ such that
\begin{equation}
	d''f_0=0,\quad hf_0=-q_0e_0 , \quad (q_0>1/2)
	\label{e01B}
\end{equation}
and define $f_n=(d')^nf_0$ ($n=1,2,...$). Then $d''f_n=b(n)f_{n-1}$ where, as follows from Eq. (\ref{e01B}), $b(0)=0,\,\,b(1)=-q_0$ and 
\begin{equation}
	b(n)=(-n-q_0+1)-b(n-1)
	\label{bn1}
\end{equation} 
The solution of this equation is 
\begin{equation}
	b(n)=-n/2-(q_0-1/2)[1-(-1)^n]/2
	\label{bnB}
\end{equation}
Therefore, $b(n) < 0\,\,\forall n$ and, as follows from Eq. (\ref{osp12}), 
$hf_n=-(n+q_0)f_n$. So, we have obtained an infinite-dimensional IR of the algebra (\ref{osp12}) with the basis $(f_0,f_1,f_2...)$ where all basis vectors are the eigenvectors of the operator $h$ with negative eigenvalues. 

\subsection{IRs of the superalgebra (\ref{osp12}) in FQT}
\label{SS2}
In FQT we can also consider an IR of the algebra (\ref{osp12}) generated by a vector $e_0$ such that
\begin{equation}
	d'e_0=0,\quad d'd''e_0=q_0e_0
	\label{e02}
\end{equation}
where now $q_0\in R_p$. As in SQT, we define $e_n=(d'')^ne_0$. 
Then $d'e_n=a(n)e_{n-1}$ where, as follows from Eq. (\ref{e02}), $a(0)=0,\,\,a(1)=q_0$
and for $a(n)$ we have the same equation as in (\ref{an1}):
\begin{equation}
	a(n)=q_0+n-1-a(n-1)
	\label{an2}
\end{equation}
We assume that $p$ is odd and then, instead of the solution (\ref{anB}), the solution is  
\begin{equation}
a(n)=\frac{p+1}{2}n+\frac{p+1}{2}(q_0-\frac{p+1}{2})[1-(-1)^n]
\label{anC}
\end{equation}
Then, unlike the situation in SQT where $a(n)>0$ at $(n=1,2,...\infty)$, we have that, since in $R_p$ the results should be taken modulo $p$, $a(n)=0$ if $n=2p+1-2q_0$. Therefore, the IR
under consideration is finite-dimensional, the basis of this IR consists of vectors
$(e_0,e_1,...e_{N})$ where $N=n_{max}=2p-2q_0$ and the dimension of the IR is $2p+1-2q_0$.

Just like in SQT, we have that $he_n=(n+q_0)e_n$ at $n=0,1,...N$, i.e., the basis vectors $e_n$ are eigenvectors of the operator $h$ with the eigenvalues $\lambda_n=n+q_0$. Since all the
$\lambda_n$ should be different in $R_p$, it should be $N\leq (p-1)$.

We choose $q_0=(p+1)/2+a$ where $a\in R_p$ and $a$ can be one of the values $(0,1,..(p-3)/2)$.
Then $f(q_0)<0$, i.e., $q_0$ is in the left half-plane of Figure \ref{Fig.2}. Then,
unlike the situation in SQT where $\lambda_n>0$ at all $n=0,1,...\infty$, we have that in FQT $f(\lambda_n)<0$ at $n=(0,1...,(p-3)/2-a)$, $f(\lambda_n)=0$ at $n=(p-1)/2-a$ 
and finally $f(\lambda_n)>0$ at $n=((p+1)/2)-a,...,n_{max})$.

Thus, the construction of the basis begins with the basis vector $e_0$ with the eigenvalue
of $h$ equal to $\lambda_0=q_0$ and ends with the basis vector $e_N$ with the eigenvalue $\lambda_N=-q_0$.
{\bf Thus, in contrast to SQT where there are two different IRs with positive and negative eigenvalues of $h$, respectively, in FQT there is only one IR which contains the analogs of negative and positive IRs in SQT and contains a basis vector with the eigenvalue of $h$ equal to zero. In addition, if in SQT, negative and positive IRs are infinite-dimensional, then in FQT, the only IR is finite-dimensional with dimension $p-2a$.}

\section{Dirac supersingleton}
\label{Dsingl}

When describing elementary particles within the framework of AdS symmetry, the following problems arise. 

If $m$ is the mass of a particle in Poincare invariant theory then its
mass $\mu$ in AdS theory is dimensionless and the relation between $\mu$ and $m$
is $\mu=mR$ where $R$ is the contraction parameter for the transition from AdS to Poincare  symmetry. As explained in \cite{asymm}, 
the data on cosmological acceleration show that, at the present stage of the universe,
$R$ is of the order of $10^{26}m$ \cite{asymm}. Therefore, even for elementary particles, the  AdS masses are very large. For example, the AdS masses of the electron,
the Earth and the Sun are of the order of $10^{39}$, $10^{93}$ and $10^{99}$, respectively. 
The fact that even the AdS mass of the electron is so large might be an
indication that the electron is not a true elementary particle. 
In addition, the present upper level for the photon mass is
$10^{-17}ev$ or less. This value
seems to be an extremely tiny quantity. However, the corresponding AdS mass is of the order of $10^{16}$ and so, even the mass which is treated as extremely small in Poincare
invariant theory might be very large in AdS invariant theory.

As shown in \cite{asymm}, in SQT, dS symmetry is more
general than AdS one but in the framework of dS symmetry it is not
possible to describe neutral elementary particles, i.e., particles which are equivalent
to the their antiparticles. As shown in Sec. \ref{FQTvsSQT}, in FQT, dS and AdS symmetries are equivalent and, as shown
in \cite{book,arxiv}, in this theory also there are no neutral elementary particles. In particular, even the photon is not elementary. 

This problem has been discussed by several authors.
In Standard Model (based on Poincare invariance) only massless particles are treated as elementary. However, as shown in the seminal paper by Flato and Fronsdal \cite{FF}
(see also \cite{HeidenreichS}), in standard AdS theory, each massless IR can be 
constructed from the tensor product of two singleton IRs discovered by Dirac in his paper \cite{DiracS} titled "A Remarkable Representation of the 3 + 2 de Sitter group", and the authors of \cite{FF}
believe that this is indeed a truly remarkable property. 

The IR describing the supersingleton is constructed as follows. In Eq. (\ref{S32A}), we choose $q_1$ and $q_2$ the same and equal $q_0$ where $q_0=1/2$ in standard theory
over complex numbers and $q_0=(p+1)/2$ in FQT, where $p$ is the characteristic of the  ring and $p$ is odd.

The authors of \cite{FF} and other publications treat singletons as true elementary 
particles because their weight diagrams has only a single trajectory (that's why
the corresponding IRs are called singletons). However, one should
answer the following questions:
\begin{itemize}
	\item a) Why singletons have not been observed yet.
	\item b) Why such massless particles as photons and others are stable and their decays into singletons have not been observed.
\end{itemize}
There exists a wide literature 
(see e.g., \cite{FFS,Bekaert} and references therein) where this problem is investigated
from the point of view of standard AdS QFT. For example, in AdS QFT, 
singleton fields live on the boundary at infinity of the AdS bulk (boundary which
has one dimension less than the bulk). 
However, as noted in Sec. \ref{FQTvsSQT}, the explanation in the framework of quantum theory should not involve classical spaces.

On the other hand, as argued in \cite{book,arxiv}, in FQT, the properties of Dirac singletons are even more remarkable than in standard theory and here the properties
a)-b) have a natural explanation.

In standard AdS theory, there exist four Dirac singletons which in the literature are called Di singleton, Rac singleton and their antiparticles. In the case of supersymmetry, Di and Rac singletons are combined into one superparticle - the Dirac supersingleton, so that there are two supersingletons - the Dirac supersingleton and its antiparticle. However, in FQT those
supersingletons are combined into one object and so there is only one supersingleton.
Here, one of the remarkable properties of supersingletons is the following. The physical meaning of division comes from classical physics, which assumes that every object can be divided into any arbitrarily large number of arbitrarily small parts. However, standard division loses its standard physical meaning when we reach the level of elementary particles since, for example, the electron cannot be divided into two, three, and so on parts. As shown in \cite{book,arxiv}, in FQT, the theory of singletons can be built 
over a ring in which there is no division, but only addition, subtraction and multiplication.

As shown in \cite{book,arxiv,FM}, the important property of supersingletons is that the above results can be immediately generalized to the case of higher dimensions, and 
in this case it is interesting to explore the possibility that  
spatial and internal quantum numbers are on equal footing. The fact that singleton physics can be directly
generalized to the case of higher dimensions has been indicated by several authors
(see e.g., \cite{FFS} and references therein). 

\section{Supersingleton IRs in SQT and FQT}
\label{DsinglIRs}
As shown in \cite{book,arxiv}, $[d_1'',d_2'']=0$ in the space of the supersingleton IR. For this reason,
the results for supersingleton IRs can be obtained by directly generalizing the results
of Subsecs. \ref{SS1} and \ref{SS2}.

\subsection{Supersingleton IRs in SQT}
\label{SingleSQT}
In this subsection it will be shown that in SQT, there is only one supersingleton IR with positive energies and 
only one supersingleton IR with negative energies.

Consider first the representation generated by a vector $e_0$ such that
\begin{equation}
	d_j'e_0=0,\quad h_je_0=\frac{1}{2}e_0 , \quad j=1,2.
	\label{sssqtplus}
\end{equation}
The basis of the IR consists of the vectors $e_{jk}=(d_1'')^j(d_2'')^ke_0$. Then, as follows from Eq. (\ref{anB})
\begin{equation}
h_1e_{jk}=(j+\frac{1}{2})e_{jk},\,\,h_2e_{jk}=(k+\frac{1}{2})e_{jk},\,\,
d_1'e_{jk}=\frac{j}{2}e_{j-1,k},\,\,d_2'e_{jk}=\frac{k}{2}e_{j,k-1}
\label{h1h2}
\end{equation}
where $j,k=0,1,2,...\infty$. As shown in \cite{book,arxiv}, $M_{04}$ is the AdS analog of the energy operator because $M_{04}$ becomes the Poincare energy upon contraction of the AdS algebra to the Poincare algebra. Then, as follows from Eqs. (\ref{Mab}) and (\ref{h1h2}),
\begin{equation}
M_{04}e_{jk}=(1+j+k)e_{jk}
\label{M04jk}
\end{equation}
and therefore, the positive energy IR is infinite-dimensional. 

Consider now the representation generated by a vector $f_0$ such that
\begin{equation}
	d_j''f_0=0,\quad h_jf_0=-\frac{1}{2}e_0 , \quad j=1,2.
	\label{sssqtminus}
\end{equation}
The basis of the IR consists of the vectors $f_{jk}=(d_1')^j(d_2')^kf_0$. Then, as follows from Eq. (\ref{bnB})
\begin{equation}
h_1f_{jk}=-(\frac{1}{2}+j)f_{jk},\,\,h_2{jk}=-(\frac{1}{2}+k)f_{jk},\,\,
d_1''f_{jk}=-\frac{j}{2}f_{j-1,k},\,\,d_2''f_{jk}=-\frac{k}{2}f_{j,k-1}
	\label{h1h2B}
\end{equation}
where $j,k=0,1,2,...\infty$. As follows from Eqs. (\ref{Mab}) and (\ref{h1h2B}),
\begin{equation}
	M_{04}f_{jk}=-(1+j+k)e_{jk}
\label{minusM04jk}
\end{equation}
and therefore, the negative energy IR is infinite-dimensional. 

\subsection{Supersingleton IRs in FQT}
In this subsection, it will be shown that, in contrast to the situation in
SQT, in FQT there exists only one supersingleton IR. For definiteness, we assume
that $p$ is odd. Then the FM analog of Eq. (\ref{sssqtplus}) is
\begin{equation}
	d_j'e_0=0,\quad h_je_0=\frac{p+1}{2}e_0 , \quad j=1,2.
	\label{ssfqtplus}
\end{equation}
and the FQT analog of Eq. (\ref{h1h2}) is
\begin{eqnarray}
&&h_1e_{jk}=(j+\frac{p+1}{2})e_{jk},\,\,h_2e_{jk}=(k+\frac{p+1}{2})e_{jk}\nonumber\\
&&d_1'e_{jk}=\frac{j(p+1)}{2}e_{j-1,k},\,\,d_2'e_{jk}=\frac{k(p+1)}{2}e_{j,k-1}
\label{FQTh1h2}
\end{eqnarray}
Since now the results should be taken modulo $p$, it follows from these expressions
that $d_1'e_{jk}=0$ at $j=p$ and $d_2'e_{jk}=0$ at $k=p$. Therefore  
$e_{jk}\neq 0$ at $j,k=0,1,..p-1$. 

{\bf We conclude that, in contrast to the situation in SQT, where there are two infinite-dimensional IRs (one positive-energy IR and one negative-energy IR), in FQT there is only one IR which is finite-dimensional with the dimension $p^2$.} 

As follows from Eq. (\ref{FQTh1h2}), since now the results should be taken modulo $p$,
we have for the eigenvalues of the operator $M_{04}$ formally the same result as
in Eq. (\ref{M04jk}) that the elements $e_{jk}$ are the eigenvectors of the operator
$M_{04}=h_1+h_2$ with the eigenvalues $(j+k+1)$. Therefore when $j,k\ll p$ we have
an analog of positive energy IR. On the other hand, if $j=p-1-j'$ and $k=p-1-k'$ then, since the results should be taken modulo $p$, we have that in terms of $j'$ and $k'$
the eigenvalues of $M_{04}$ are equal to $-(1+j'+k')$ by analogy with Eq. (\ref{minusM04jk}). Therefore when $j',k'\ll p$ we have an analog of negative energy IR.

\section{Conclusion}

In this paper we note that, as shown in our previous works, fundamental quantum theory should be based on finite mathematics in which it is not assumed
that the characteristic $p$ of the ring used in this mathematics is anomalously large.
In this theory there are no concepts of particle-antiparticle and conserved additive quantum numbers such as electric charge, baryon quantum number etc.

The above properties have been discussed in our publications with detailed technical proofs. The purpose of this paper is to consider models where differences between
SQT and FQT can be described in a much simpler way. In Sec. \ref{model} we consider a model where a superalgebra is defined by only two operators and in Sec.
\ref{DsinglIRs} we consider the model of Dirac supersingleton.

Our publications and this paper illustrate that the standard concepts of particle-antiparticle and conserved additive quantum numbers are not fundamental. As noted in \cite{book,arxiv,lev3,FM}, our results on universal law of gravity indicate that at the present state of the universe the value $p$ is very large
(of the order of $exp(10^{80})$ or more) and that is why these concepts work approximately with very high accuracy. However, there is no reason to think that $p$ is a fundamental quantity that had the same value at all stages of the universe.

Each computer can carry out calculations only modulo a certain number, which depends on the maximum number of bits with which this computer can work. The literature discusses the possibility that the universe can be treated as a computer (see e.g., \cite{Wolfram}).
From this point of view, the value $p$ is not some fundamental constant but
is determined by the state of the universe at a given stage. 
And, since the state of the universe is changing, it is natural to expect that
the number $p$ describing physics at different stages of the evolution of the universe
will be different at different stages.

There are several reasons to think that at early stages of the universe the value $p$
was much less than now. One of the reasons is the problem of the baryon asymmetry of the universe (BAU). Modern cosmological theories state that the numbers of baryons and antibaryons in
the early stages of the universe were the same. Then, since the baryon number is the
conserved quantum number, those numbers should be the same at the present stage.
However, at this stage the number of baryons is much greater than the number of
antibaryons. However, if the value $p$ at early stages of the universe was much less than now then the statement that the numbers of baryons and antibaryons were the same, does not have a physical meaning and the BAU problem does not arise (see e.g., \cite{book,arxiv,FM} for more details).

Another reason is the problem of time in quantum theory. As noted in Sec. \ref{background},
this problem was posed by Pauli in view of the fact that in this theory there is no time operator. In \cite{book,arxiv} we discussed a conjecture that standard classical time $t$ manifests itself because the value $p$ changes, i.e., $t$ is a function of $p$. We do not say that $p$ changes over time because classical time $t$ cannot be present in quantum theory; we say that we feel changing time because $p$ changes. In \cite{time} we discussed
a model where, in semiclassical approximation, the variations of $t$ and $p$ at the present stage of the universe, are related as 
$$\Delta t=\frac{R}{c}\frac{\Delta ln(p)}{ln(p)}$$
where $R$ is the contraction parameter from the dS to the Poincare algebra. In this model, the quantities $t$ and $p$ increase during the evolution of the universe. However, since we do not know how the states of the universe were described in its early stages, we cannot say what the magnitudes of $p$ were at these stages.

In summary, the concepts of particle-antiparticle and additive quantum numbers 
should not be present in the ultimate quantum theory which should not assume 
that $p$ is necessarily very large. Thus, the main problem is what principles this theory should be based on. 

{\bf Acknowledgements:} I am grateful to Vladimir Karmanov and to the reviewers of this paper for important remarks. 

{\bf Data Availability Statement}: No new data were created or analyzed in this study. Data sharing is not applicable to this article.

{\bf Funding}: This research received no external funding.

{\bf Conflicts of Interest}: The author declares no conflicts of interest. 


\end{document}